\documentclass[final,5p,times,twocolumn]{elsarticle}
\usepackage{graphics}

\begin{document}

\begin{frontmatter}

\title{Anisotropic magnetization and resistivity of single crystalline \textit{R}Ni$_{1-x}$Bi$_{2 \pm y}$ (\textit{R} = La--Nd, Sm, Gd--Dy)}

\author[label1]{Xiao Lin} \author[label2]{Warren E. Straszheim} \author[label1,label2]{Sergey L. Bud'ko} \author[label1,label2]{Paul C. Canfield}

\address[label1]{Department of Physics and Astronomy, Iowa State University, Ames, Iowa 50011, U.S.A.}

\address[label2]{Ames Laboratory, US DOE, Iowa State University, Ames, Iowa 50011, U.S.A}

\begin{abstract}

We present a detailed study of \textit{R}Ni$_{1-x}$Bi$_{2 \pm y}$ (\textit{R} = La--Nd, Sm, Gd--Dy) single crystals by measurements of stoichiometry and temperature dependent magnetic susceptibility, magnetization, and electrical resistivity. This series forms with partial Ni occupancy as well as a variable Bi occupancy. For \textit{R} = Ce--Nd, Gd--Dy, the \textit{R}Ni$_{1-x}$Bi$_{2 \pm y}$ compounds show local-moment like behavior and order antiferromagnetically at low temperatures. Determination of anisotropies as well as antiferromagnetic ordering temperatures for \textit{R}Ni$_{1-x}$Bi$_{2 \pm y}$ (\textit{R} = Ce--Nd, Sm, Gd--Dy) have been made. Although crystalline samples from this family exhibit minority, second phase superconductivity at low temperatures associated with Ni-Bi and Bi contamination, no evidence of bulk superconductivity has been observed.

\end{abstract}

\begin{keyword}
Rare-earth compounds \sep Single crystals \sep Magnetization \sep Resistivity 

\end{keyword}

\end{frontmatter}

\section{Introduction}

The interesting physical properties of Ce-based intermetallic compounds have been the concern of numerous studies \cite{CeCu2Si2, Myers_1999, Sergey_1999, Petrovic_2003}. The ground state of Ce-based intermetallic compounds is often governed by the competition between the Ruderman-Kittel-Kasuya-Yosida (RKKY) interaction and the Kondo interaction. Depending on the strength of the hybridization between 4\textit{f} and conduction electrons relative to their coupling strength, the ground state can be either a non-magnetic state dominated by the Kondo interaction or a long-range magnetically ordered state governed by the RKKY interaction. Various exotic phenomena have been observed in these compounds, for example: CeCu$_2$Si$_2$ was classified as a heavy fermion superconductor \cite{CeCu2Si2}, CeAgSb$_2$ \cite{Myers_1999} and CeAgBi$_2$ \cite{Petrovic_2003} were reported to be strongly correlated electron compounds with a magnetically ordered ground state.

It is often enlightening to study not just the Ce-member of a rare earth, intermetallic series, but rather a wide sampling of the whole series. As exhibited in many rare earth series \cite{Myers_1999, Sergey_1999, Petrovic_2003, Handbook1, Handbook2}, the 4\textit{f} electrons are often shielded from the 5\textit{s}-, 5\textit{p}- and 4\textit{d}-shell electrons, and thus do not participate in chemical bonding. On the other hand, the 4\textit{f} electrons have direct influences on the compounds' magnetic properties, since the magnetic moments originate in the partially filled \textit{f}-shell. As a result, by varying the \textit{R} elements in a compound, it is possible to tune the magnetism and other physical properties. Moreover, the unit cell volume of isostructural $R^{3+}$-bearing families shrinks from \textit{R} = La to \textit{R} = Lu, which is known as the lanthanide contraction. This contraction leads to systematic changes in the lattice constants \textit{a}, \textit{b}, \textit{c} and unit cell volume \textit{V}.

Recently CeNi$_{0.8}$Bi$_2$ has attracted particular attention: a polycrystalline sample was synthesized and reported as a heavy fermion superconductor \cite{Hosono1}. Its $T_{\rm c}$ was found to be $\sim$ 4.2 K. The superconductivity was said to be associated with the light effective mass electrons, whereas the antiferromagnetic (AFM) transition, occurring at $\sim$ 5 K, was related to the strong interactions between the heavy electrons and Ce 4\textit{f} electrons \cite{Hosono1}. The superconductivity was claimed to be introduced by the Ni deficiency, as the ``parent" compound CeNiBi$_2$ did not manifest bulk superconductivity \cite{Hosono2}. Earlier results on CeNiBi$_2$ suggested it to be a moderately heavy fermion antiferromagnet \cite{Onuki_Ce, Takabatake_Ce}. Magnetic susceptibility was measured on single crystalline samples with no diamagnetic signal being observed, and the zero resistivity was attributed ``to the thin films of bismuth" \cite{Onuki_Ce}. 

Given that Ni occupancy in the sample is claimed to play a key role in superconductivity, single crystalline samples of CeNi$_{0.8}$Bi$_2$ are likely to offer further understandings of the superconducting features in this system. Hence, we present a study of physical properties of single crystalline CeNi$_{0.8}$Bi$_2$ samples.

The early work on the polycrystalline \textit{R}Ni$_{1-x}$Bi$_{2 \pm y}$ samples (\textit{R} = Ce, Nd, Gd, Tb, Dy and Y) solved the structure and reported the Ni site deficiency \cite{Franzen, Zhao}. This series of compounds was found to have a tetragonal ZrCuSi$_2$-type structure (space group \textit{P4/nmm}). Later superconductivity with $T_{\rm c} \sim$ 4 K was reported in this family for \textit{R} = Y, La, Ce and Nd \cite{Hosono1, Hosono3}. The results were based on the polycrystalline samples containing partial occupancy of the Ni site. Other than for CeNi$_{0.8}$Bi$_2$, no detailed, anisotropic results were shown for other members in this family.

In this paper, we present a systematic study of the anisotropic properties of the \textit{R}Ni$_{1-x}$Bi$_{2 \pm y}$ series with \textit{R} = La--Nd, Sm, Gd--Dy. Since this system shows partial occupancy of the Ni site, chemical elemental analysis was performed to determine the stoichiometry of the samples. Analyses of the field and temperature dependence of the magnetization and resistivity were performed on the single crystalline samples. No evidence of bulk superconductivity was observed in this series of compounds. For \textit{R} = Ce--Nd, Gd--Dy, compounds show local-moment like behavior with an AFM ordering at low temperatures. Measurements of the magnetization parallel to the \textit{ab}-plane and the \textit{c}-axis show anisotropic behavior, and the magnetization of some of the compounds indicate the existence of metamagnetic transitions. 

\section{Experimental details}
Single crystals of \textit{R}Ni$_{1-x}$Bi$_{2 \pm y}$ were grown out of excess Bi flux via the high-temperature solution method \cite{Canfield_1992, Canfield_2010}. Given that this series of compounds is known to have Ni site deficiency, and that the Ni deficiency was considered to be crucial for CeNi$_{1-x}$Bi$_{2}$ to become superconducting \cite{Hosono1}, we varied the composition of the starting materials, resulting in crystals with different Ni concentrations. Single crystals of CeNi$_{1-x}$Bi$_{2}$ with (1-\textit{x}) varying from 0.64 to 0.85 (as determined from wavelength-dispersive x-ray spectroscopy, see below) were synthesized. Irreproducible and incomplete transitions in resistivity data were seen for (1-\textit{x}) = 0.64, 0.75, 0.80 and 0.85. They are very likely associated with minority, second phase superconductivity. No qualitative differences were observed for these samples with different (1-\textit{x}) values. In this work the starting stoichiometry that gives the resulting crystal with the ratio of Ce:Ni:Bi = 1:0.8:2 was selected. This initial stoichiometry was Ce$_{10.4}$Ni$_{14.6}$Bi$_{75}$. The same initial stoichiometry was used for \textit{R} = La, Pr, Nd and Sm. For \textit{R} = Gd, Tb and Dy, to avoid \textit{R}Bi as an impurity, the stoichiometry was adjusted to \textit{R}$_{4.5}$Ni$_{9.1}$Bi$_{86.4}$. For \textit{R} = Eu and Ho -- Lu, there are no reported data on the isostructural compounds. Our attempts to grow these \textit{R}-members of the series often resulted in poorly-formed \textit{R}-Bi binaries. Hence, it is likely that this series does not form under the similar growth conditions for \textit{R} = Eu and Ho -- Lu. 

High purity ($>$3N) elements were placed in an alumina crucible and sealed in a fused silica tube under a partial pressure of high purity argon gas. This was then heated up to 1000$^\circ$C and slowly cooled to 500$^\circ$C, at which temperature the excess solution was decanted using a centrifuge \cite{Canfield_1992, Canfield_2010}. Single crystals of \textit{R}Ni$_{1-x}$Bi$_{2 \pm y}$ grew in plate-like shapes with their sizes varying from $\sim$ $10 \times 10 \times 1.5$ mm$^3$ for \textit{R} = La to $\sim$ $2 \times 2 \times 0.6$ mm$^3$ for \textit{R} = Dy. The crystallographic \textit{c}-axis is perpendicular to the plate-like single crystals. Most of the samples had shiny surfaces that were partially covered by secondary phase materials. Due to the samples' air-sensitivity, crystals were kept in an argon glove-box, and efforts were made to minimize their exposure to air during samples' manipulation and measurements. They were neither etched nor polished, and only cleaved samples with fresh surface were used in the resistivity measurements. 

Powder x-ray diffraction data were collected on a Rigaku MiniFlex diffractometer with Cu K$\alpha$ radiation at room temperature. The sample was ground in a glove-box and the powder was protected from atmosphere by Kapton film during the measurement so as to protect it from oxidation. The error bars associated with the values of the lattice parameters were determined by statistical errors, and Si powder standard was used as the internal reference.

Elemental analysis of the samples was performed using wavelength-dispersive x-ray spectroscopy (WDS) in a JEOL JXA-8200 electron probe microanalyzer. Only clear and shiny surface regions were selected for determination of the sample stoichiometry, i.e. regions with residual Bi flux were avoided. For each compound, the WDS data were collected from multiple points on the same sample.

Measurements of field dependent magnetization and temperature dependent susceptibility were performed in a Quantum Design, Magnetic Property Measurement System (MPMS). The ac resistivity was measured by a standard four-probe method in a Quantum Design, Physical Property Measurement System (PPMS) or with LR700 ac resistance bridge in MPMS. Platinum wires were attached to the sample using Dupont 4929 silver paint  with the current flowing in the \textit{ab}-plane. The absolute values of resistivity are accurate to $\pm$ 20$\%$ due to the irregularity of the sample geometry and accuracy of measurements of electrical contacts' position. The residual resistivity ratio is determined as (RRR) = $\rho$(300 K) / $\rho$(5.0 K), so as to avoid any contamination from the minority phase superconductivity.  

\section{Results and analysis}
\subsection{Crystal stoichiometry and structure}

The stoichiometry of the \textit{R}Ni$_{1-x}$Bi$_{2 \pm y}$ samples was inferred from WDS analyses. Table \ref{table:WDS} summarizes the atomic percent of each element determined from the weight percent obtained from the analysis. The precision of the analysis was calculated by \textit{SD}/$\sqrt{N}$, where \textit{SD} is the standard deviation of measurements, and \textit{N} is the number of points taken in analysis. A higher level of oxygen contamination was detected in DyNi$_{0.74}$Bi$_{1.76}$ compound, which results in a lower accuracy of its analysis. The averaged atomic concentrations of each element in each compound were normalized to the rare earth element to have $R_{1.00}$. The result shows that although the ratio of \textit{R}:Ni:Bi is grossly 1:1:2, a significant Ni deficiency develops across the series, and the Bi concentration varies from slight excess to slight deficiency. Given this series of compounds does not maintain a fixed, stoichiometric composition, for the calculation of physical quantities, the stoichiometries from Table \ref{table:WDS} were used. It should be noted that if the Bi stoichiometry is simply set to 2.00, the values for effective moments vary significantly as the molar mass changes. For this reason the Bi nonstoichiometry is also shown.

\begin{figure}
\begin{center}
\resizebox*{7.5cm}{!}{\includegraphics{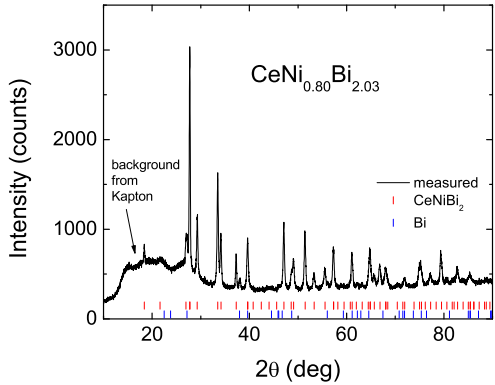}}%
\caption{ Powder x-ray diffraction pattern of CeNi$_{0.80}$Bi$_{2.03}$.}%
\label{fig:x-ray}
\end{center}
\end{figure}

\begin{figure}
\begin{center}
\resizebox*{7.5cm}{!}{\includegraphics{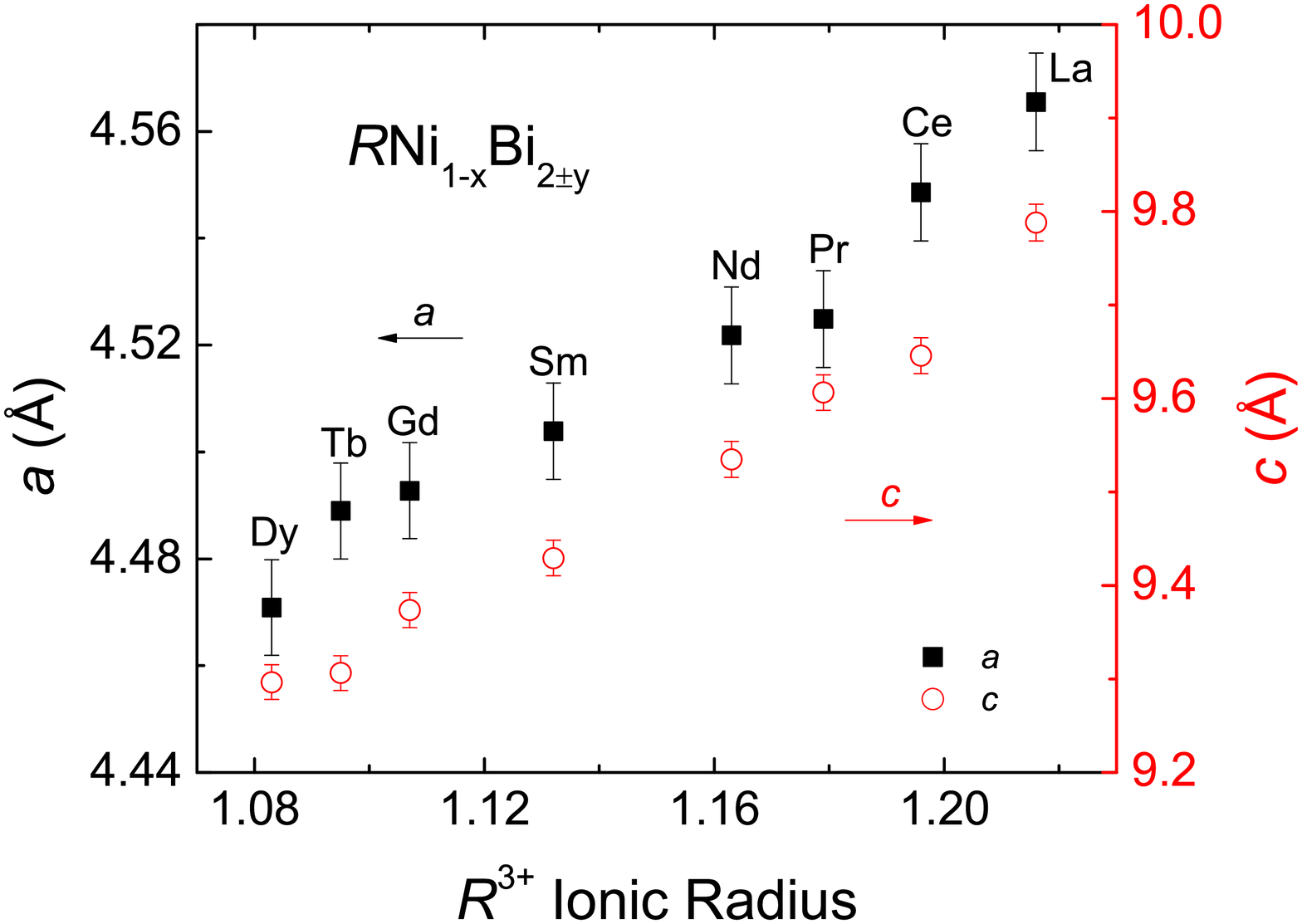}}%
\caption{The change of unit cell lattice parameters vs. ionic radius of $ R^{3+}$ for 9 coordination number (CN=9) \cite{Shannon_1976} in \textit{R}Ni$_{1-x}$Bi$_{2 \pm y}$ compounds.}%
\label{fig:lattice}
\end{center}
\end{figure}

Powder x-ray diffraction patterns were collected on ground single crystals from each compound. Figure \ref{fig:x-ray} shows a CeNi$_{0.80}$Bi$_{2.03}$ x-ray pattern as an example. The main phase was fitted with CeNiBi$_{2}$'s diffraction pattern (site occupancy was not analysed), small traces of Bi residue can be detected, whereas no evidence of Ni-Bi binaries was found. Similar results (\textit{R}NiBi$_{2}$ with minority phase of Bi) were obtained for the other members of the series. The analysis of powder x-ray diffraction data indicates that the lattice parameters \textit{a} and \textit{c} are monotonically decreasing as the series progresses from La to Dy (presented in Table \ref{table:powder}). Proceeding from the larger to the smaller rare-earth elements, all lattice parameters decrease almost linearly: ~2.1$\%$ for \textit{a} and ~5.3$\%$ for \textit{c} (as shown in Fig. \ref{fig:lattice}), which is consistent with the previously reported data \cite{ Hosono1, Franzen, Zhao}. In progressing from LaNi$_{0.84}$Bi$_{2.04}$ to DyNi$_{0.74}$Bi$_{1.76}$, the overall volume decreases by ~9.8$\%$. These results are likely associated with the lanthanide contraction that occurs across the 4\textit{f} series. In addition, the deficiencies at the Ni site can possibly lead to smaller unit cell volumes as well \cite{Mun_2010}.

\subsection{Resistivity of \textit{R}Ni$_{1-x}$Bi$_{2 \pm y}$}
\begin{figure}
\begin{center}
\resizebox*{7.5cm}{!}{\includegraphics{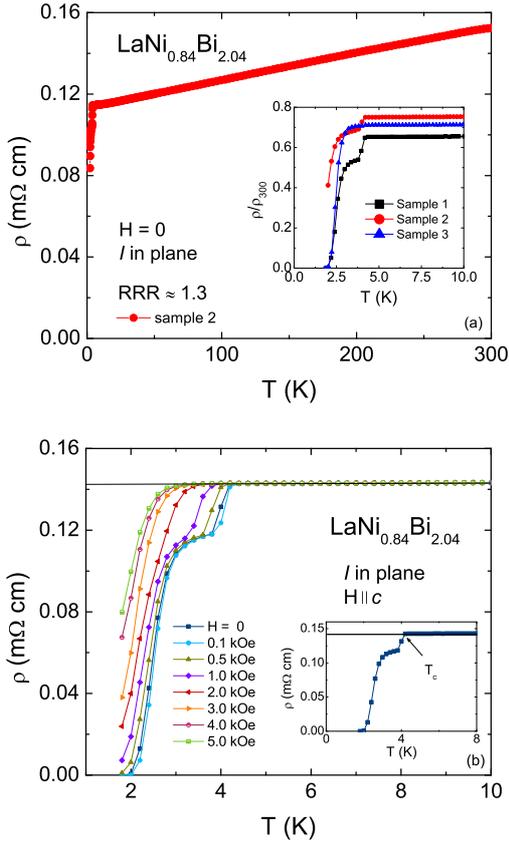}}%
\caption{ (a) Temperature dependence of the electrical resistivity of LaNi$_{0.84}$Bi$_{2.04}$. Inset: comparison of the enlarged normalized resistivity ratio from three samples for $T \leq$ 10 K. (b) Low temperature resistivity of LaNi$_{0.84}$Bi$_{2.04}$ measured at 0, 0.1, 0.5, 1.0, 2.0, 3.0, 4.0, and 5.0 kOe with \textbf{H} $\parallel c$. Inset: criteria for the higher-temperature transition $T_{\rm c}$ is shown for \textit{H} = 0.}%
\label{fig:La_RT}
\end{center}
\end{figure}

The temperature dependent electrical resistivity for LaNi$_{0.84}$Bi$_{2.04}$ is shown in Fig. \ref{fig:La_RT} (a). To within a factor of 20$\%$, the room temperature resistivity value $\rho$(300 K) reaches $\sim$ 0.15 m$\Omega$ cm. For current in the \textit{ab}-plane, the resistivity displays metallic behavior with RRR $\approx$ 1.3. We assume that this small RRR value is at least partially due to the deficiencies at the Ni and Bi sites. Although the high temperature ($T >$ 5 K) resistivity is quite reproducible, sample to sample, below 5 K the sizes of the two resistive anomalies (observed near 4.2 K and 3.0 K) are very sample dependent (Fig. \ref{fig:La_RT} (a) inset). Whereas the resistivity of Sample 2 does not reach zero above \textit{T} = 1.8 K, the higher-temperature anomaly is barely seen for Sample 3. Such incomplete and irreproducible transitions suggest that the superconductivity is extrinsic in both cases and the anomalies in the resistivity can be attributed to minority phases. In fact, it is highly likely that the higher-temperature anomaly is related to the superconducting transition of the Ni-Bi binaries (NiBi with $T_{\rm c} \approx$ 4.25 K and NiBi$_{3}$ with $T_{\rm c} \approx$ 4.06 K)\cite{Ni-Bi, Petrovic_2012}. The lower-temperature anomaly is probably caused by filamentary, thin film, Bi presented in the sample, which is consistent with the previous work \cite{Onuki_Ce}; the $T_{\rm c}$ of Bi film can vary from 2 K to 5 K depending on its thickness \cite{Bi}. Moreover, the drop of resistivity at $\sim$ 4.2 K is always much smaller then the drop at $\sim$ 3.0 K. This is possibly due to smaller amount of Ni-Bi binaries than Bi film in the sample, which is consistent with the x-ray diffraction data (where Bi is clearly detected, and no Ni-Bi binary diffraction lines can be resolved). The low temperature resistivity data, measured in different magnetic fields, are plotted in Fig. \ref{fig:La_RT} (b). As can be seen, both of the superconducting features are shifting to lower temperatures as the applied field increases. 

\begin{figure}
\begin{center}
\resizebox*{7.5cm}{!}{\includegraphics{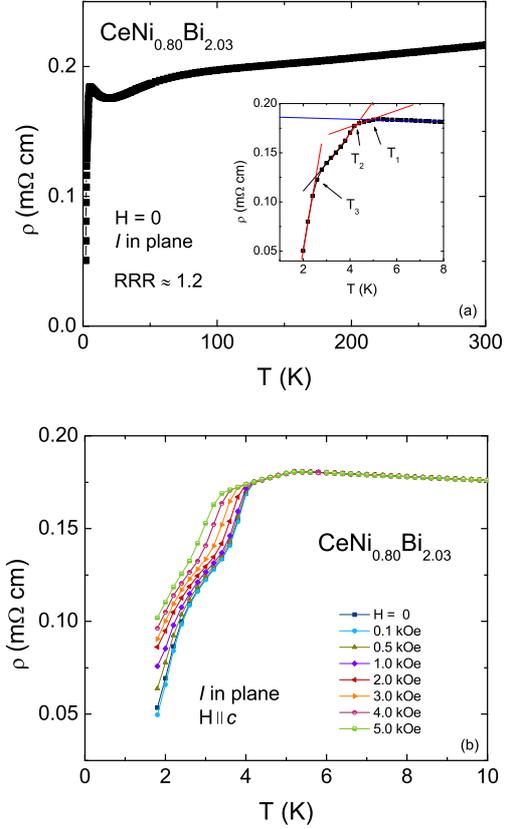}}%
\caption{ (a) Temperature dependence of the electrical resistivity of CeNi$_{0.80}$Bi$_{2.03}$. Inset: criterion for the transition temperature $T_{\rm 1}$, $T_{\rm 2}$ and $T_{\rm 3}$ are shown for \textit{H} = 0. (b) Low temperature resistivity of CeNi$_{0.80}$Bi$_{2.03}$ measured at 0, 0.1, 0.5, 1.0, 2.0, 3.0, 4.0, and 5.0 kOe with \textbf{H} $\parallel c$. }%
\label{fig:Ce_RT}
\end{center}
\end{figure}

The temperature dependent electrical resistivity for CeNi$_{0.80}$Bi$_{2.03}$ is shown in Fig. \ref{fig:Ce_RT} (a). To within a factor of 20$\%$, the room temperature resistivity value $\rho$(300 K) reaches $\sim$ 0.22 m$\Omega$ cm with RRR $\approx$ 1.2. A broad hump at $\sim$ 80 K and a local minimum at $\sim$ 20 K are found. These features are readily associated with the interplay of the thermal population of the crystal electric field (CEF) levels and the Kondo effect, which are often seen in the Ce-based compounds \cite{Myers_1999, Petrovic_2003}. At low temperatures, three successive resistive anomalies are detected near 5.1 K, 4.2 and 2.5 K (see inset of Fig. \ref{fig:Ce_RT} (a)). The criteria for determining the transition temperatures are shown in the inset of Fig. \ref{fig:Ce_RT} (a). The highest transition temperature ($T_{\rm 1} \approx$ 5.1 K) coincides with the AFM transition, which will be shown in the magnetic susceptibility data discussed in the next section. It is likely caused by the loss of the spin-disorder scattering upon entering into the magnetically ordered state. Since the resistivity does not reach zero above 1.8 K, it implies that CeNi$_{0.80}$Bi$_{2.03}$ does not show bulk superconducting behavior, and the two lower-temperature anomalies are probably associated with minority second phase of Ni-Bi binaries and Bi present in the sample. The low temperature resistivity data measured in different magnetic fields are plotted in Fig. \ref{fig:Ce_RT} (b). As can be seen, whereas the highest transition, $T_{\rm 1}$, is almost invariant under these fields, whereas both $T_{\rm 2}$ and $T_{\rm 3}$ decrease as the applied field increases.

\begin{figure}
\begin{center}
\resizebox*{7.5cm}{!}{\includegraphics{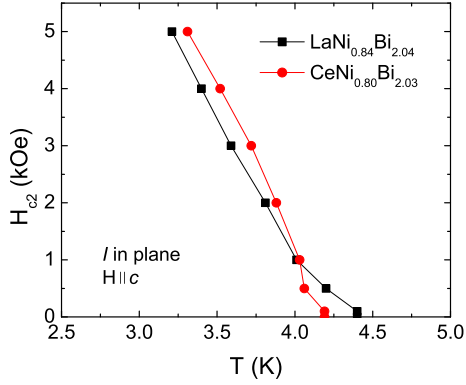}}%
\caption{ $H_{\rm c2} (T)$ plot for the higher-temperature transition of LaNi$_{0.84}$Bi$_{2.04}$ and $T_{\rm 2}$ transition of CeNi$_{0.80}$Bi$_{2.03}$ from the magnetotransport measurements.}%
\label{fig:HT}
\end{center}
\end{figure}

The upper critical fields for the superconductivity associated with the second phase in LaNi$_{0.84}$Bi$_{2.04}$ and CeNi$_{0.80}$Bi$_{2.03}$ have been obtained from the magnetotransport data (Fig. \ref{fig:HT}). To estimated the $H_{\rm c2} (T)$ values of the higher-temperature feature of LaNi$_{0.84}$Bi$_{2.04}$, the first data point deviated from the normal state is chosen as the criterion (shown in the inset of Fig. \ref{fig:La_RT} (b)). The criterion for determining $T_{\rm 2}$ of CeNi$_{0.80}$Bi$_{2.03}$ is illustrated in the inset of Fig. \ref{fig:Ce_RT} (a). The resulting $H_{\rm c2} (T)$ curves are plotted in Fig. \ref{fig:HT}. As can be seen, the two $H_{\rm c2} (T)$ curves are essentially the same. In addition, they are also quite consistent with the reported $H_{\rm c2} (T)$ phase diagram for NiBi$_{3}$ \cite{Petrovic_2012}. The fact that $H_{\rm c2}$ values in this work are larger is probably due to the difference of the $H_{\rm c2}$ criteria and difference between single crystalline and polycrystalline samples. Hence, it is very likely that the superconducting features are not of bulk properties of LaNi$_{0.84}$Bi$_{2.04}$ or CeNi$_{0.80}$Bi$_{2.03}$, but most likely due to Bi and Ni-Bi binary impurities.

The two resistive anomalies (the higher-temperature one occurring $\sim$ 4.2 K and the lower-temperature one varying from 2 to 3 K) are detected in other members of the \textit{R}Ni$_{1-x}$Bi$_{2 \pm y}$ series (\textit{R} = Pr, Nd, Sm, Gd--Dy) as well. Similar to LaNi$_{0.84}$Bi$_{2.04}$ and CeNi$_{0.80}$Bi$_{2.03}$, the superconductivity associated with these two features are partial and very sample dependent. It is most likely that these two anomalies are related to the superconducting transitions of Ni-Bi binaries and films of Bi as well. No evidence of bulk superconductivity has been found in this family from the electrical resistivity data. Unfortunately, the two anomalies often manifest as stronger features than those related to the magnetic transitions, such as in the case of CeNi$_{0.80}$Bi$_{2.03}$. Hence, the transport data for the rest of the series are not shown in this work. 

\subsection{Magnetic properties of \textit{R}Ni$_{1-x}$Bi$_{2 \pm y}$}

\begin{figure}
\begin{center}
\resizebox*{7.5cm}{!}{\includegraphics{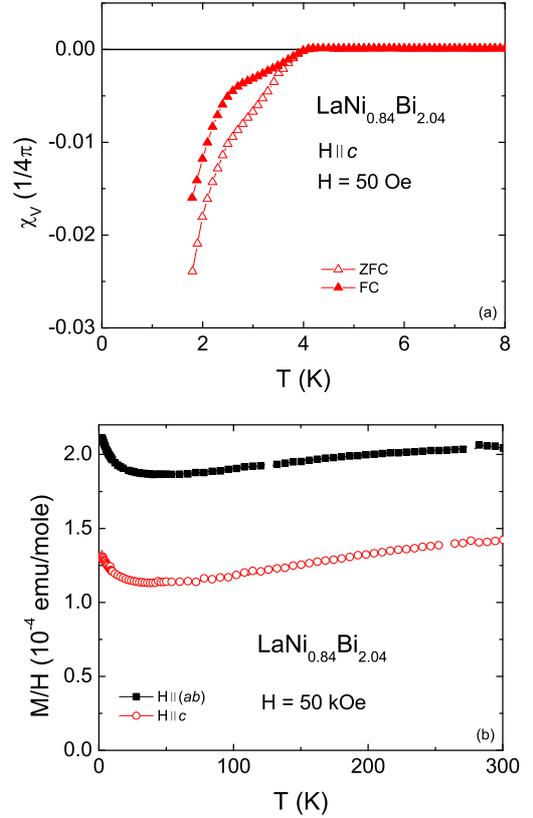}}%
\caption{ (a) The zero-field-cooled and field-cooled magnetic susceptibility of LaNi$_{0.84}$Bi$_{2.04}$ for $T \leq $ 8 K. (b) The anisotropic temperature-dependent magnetic susceptibility of LaNi$_{0.84}$Bi$_{2.04}$ for \textbf{H} $\parallel$ \textit{ab}-plane, and \textit{c}-axis.}%
\label{fig:La_MT}
\end{center}
\end{figure}

\begin{figure}
\begin{center}
\resizebox*{7.5cm}{!}{\includegraphics{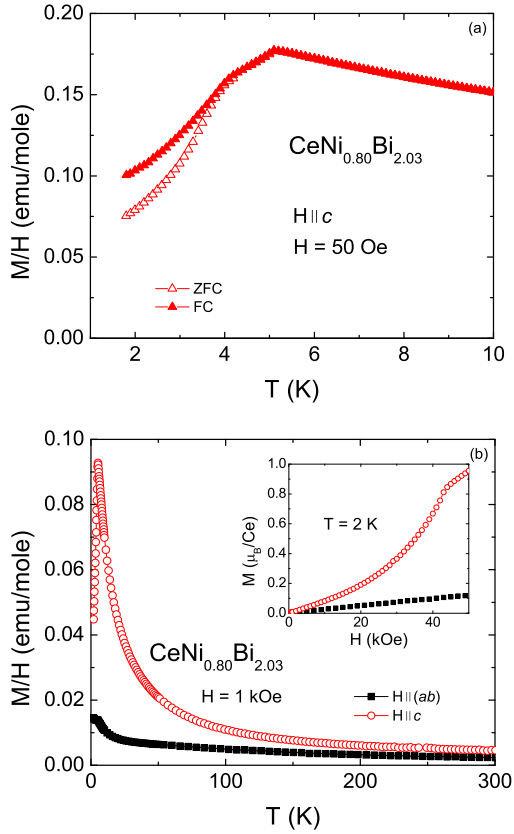}}%
\caption{ (a) The zero-field-cooled and field-cooled magnetic susceptibility of CeNi$_{0.80}$Bi$_{2.03}$ for $T \leq $ 10 K. (b) The anisotropic temperature-dependent magnetic susceptibility of CeNi$_{0.80}$Bi$_{2.03}$ for \textbf{H} = 1 kOe. Inset: the anisotropic magnetic isotherms of CeNi$_{0.80}$Bi$_{2.03}$ at \textit{T} = 2 K.}%
\label{fig:Ce_MT}
\end{center}
\end{figure}

The low temperature magnetic susceptibility of LaNi$_{0.84}$Bi$_{2.04}$ for $T \leq $ 8 K is shown in Fig. \ref{fig:La_MT} (a). With \textit{H} = 50 Oe parallel to the \textit{c}-axis, both of the zero-field-cooled (ZFC) and field-cooled (FC) magnetic susceptibilities reveal abrupt drops at around 4.1 K and 2.5 K. This is consistent with the anomalies found in the resistivity data. The small superconducting volume fractions in both ZFC ($< 3\%$) and FC ($< 2\%$) data at 2.0 K further support that LaNi$_{0.84}$Bi$_{2.04}$ does not manifest bulk superconductivity.

The anisotropic temperature-dependent magnetic susceptibility of LaNi$_{0.84}$Bi$_{2.04}$ measured from 2 K to 300 K is shown in Fig. \ref{fig:La_MT} (b). For \textit{H} = 50 kOe, LaNi$_{0.84}$Bi$_{2.04}$ shows weak paramagnetism for both orientations. With a larger value for \textbf{H} $\parallel$ \textit{ab}-plane than \textbf{H} $\parallel$ \textit{c}-axis, the magnetic susceptibility shows very subtle change as temperature decreases from 300 K. The small up-turn at low temperatures, seen in both orientations, is probably caused by small levels of paramagnetic impurities.

The low temperature magnetic susceptibility of CeNi$_{0.80}$Bi$_{2.03}$ for $T \leq $ 10 K is shown in Fig. \ref{fig:Ce_MT} (a). With \textit{H} = 50 Oe parallel to the \textit{c}-axis, CeNi$_{0.80}$Bi$_{2.03}$ shows positive susceptibility in both ZFC and FC measurements. Two sudden breaks, occurring at $\sim$ 5.1 K and 4.0 K can be seen, which are consistent with the anomalies observed in the resistivity data. No anomaly is detected for temperatures between 2.0 and 4.0 K. Exhibiting positive magnetic susceptibility together with the missing feature at low temperatures strongly support that the superconductivity observed in the resistivity data of CeNi$_{0.80}$Bi$_{2.03}$ is related to minority phases. 

The anisotropic temperature-dependent magnetic susceptibility $\chi(T)=M(T)/H$ of CeNi$_{0.80}$Bi$_{2.03}$ measured with \textit{H} = 1 kOe applied both parallel to the \textit{ab}-plane and the \textit{c}-axis are plotted in Fig. \ref{fig:Ce_MT}(b). $\chi_c$ is significantly larger than $\chi_{ab}$ over the whole temperature range measured. The sharp peaks seen at low temperature suggest that this material has an AFM transition. The ordering temperature, consistent with the reported value \cite{Onuki_Ce, Takabatake_Ce}, was estimated to be $\sim$ 4.8 K (here and in Table \ref{table:Temperature} the values of the magnetic ordering temperatures obtained from the maximum of the derivatives d($\chi$\textit{T})/d$T$ \cite{Fisher_1962} are quoted). The polycrystalline averaged susceptibility was estimated by $\chi_{ave}$ = $\frac{1}{3}$ $(\chi_{c}+2\chi_{ab})$. A modified Curie-Weiss law with inclusion of a temperature-independent term $\chi_{0}$: $\chi=\chi_{0}+\frac{C}{T-\theta}$, was used to fit the magnetic susceptibility, where \textit{C} is the Curie constant and $\theta$ is the paramagnetic Curie temperature. Considering the presence of impurities, Bi nonstoichiometry and accuracy of measuring sample's mass, the values of the effective moments in this series are accurate to $\pm 10 \%$. For CeNi$_{0.80}$Bi$_{2.03}$, it gives $\theta_{ab}$ = -156 K, $\theta_{c}$ = -6 K and $\theta_{ave}$ = -17 K, suggesting the presence of CEF splitting and AFM interaction. The inferred effective moment from the polycrystalline averaged data: $\mu_{eff}$ = 2.4 $\mu_B$/Ce is consistent with the expected Hund's rule (\textit{J} = 5/2) ground-state value, 2.54 $\mu_B$. The anisotropic field-dependent magnetization isotherms of CeNi$_{0.80}$Bi$_{2.03}$ measured at 2 K are shown in the inset of Fig.\ref{fig:Ce_MT} (b). For both orientations, $M(H)$ increases as the applied field increases. The magnetization is found to be very anisotropic with $M_c > M_{ab}$, and a cusp at $\sim$ 43 kOe for \textbf{H} $\parallel$ \textit{c} is most likely associated with a metamagnetic transition. 

\begin{figure*}
\begin{center}
\scalebox{0.6}{{\includegraphics{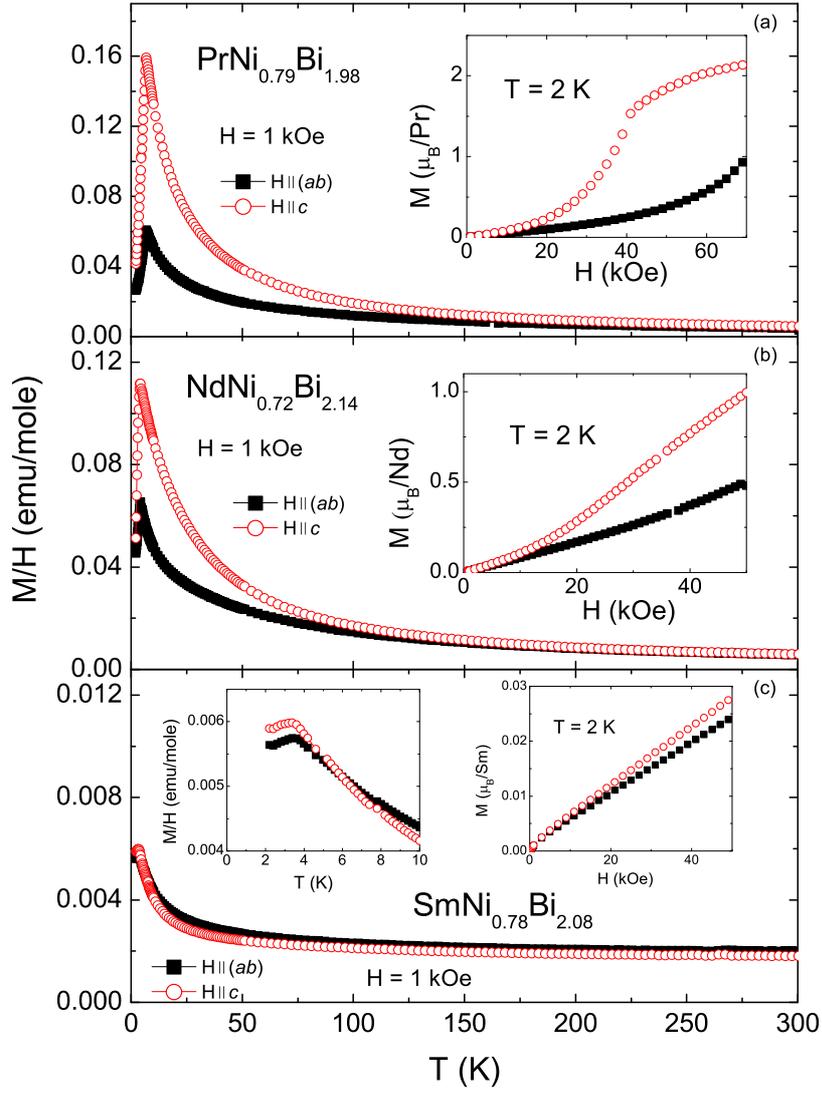}}}
\caption{(a)The anisotropic temperature-dependent magnetic susceptibility of PrNi$_{0.79}$Bi$_{1.98}$. Inset: the anisotropic magnetic isotherms of PrNi$_{0.79}$Bi$_{1.98}$.
(b) The anisotropic temperature-dependent magnetic susceptibility of NdNi$_{0.72}$Bi$_{2.14}$. Inset: the anisotropic magnetic isotherms of NdNi$_{0.72}$Bi$_{2.14}$. 
(c) The anisotropic temperature-dependent magnetic susceptibility of SmNi$_{0.78}$Bi$_{2.08}$. Left inset: enlarged magnetic susceptibility for $T \leq $ 10 K. Right inset: the anisotropic magnetic isotherms of SmNi$_{0.78}$Bi$_{2.08}$.} 
\label{fig:M_all1}
\end{center}
\end{figure*}

\begin{figure*}
\begin{center}
\scalebox{0.6}{{\includegraphics{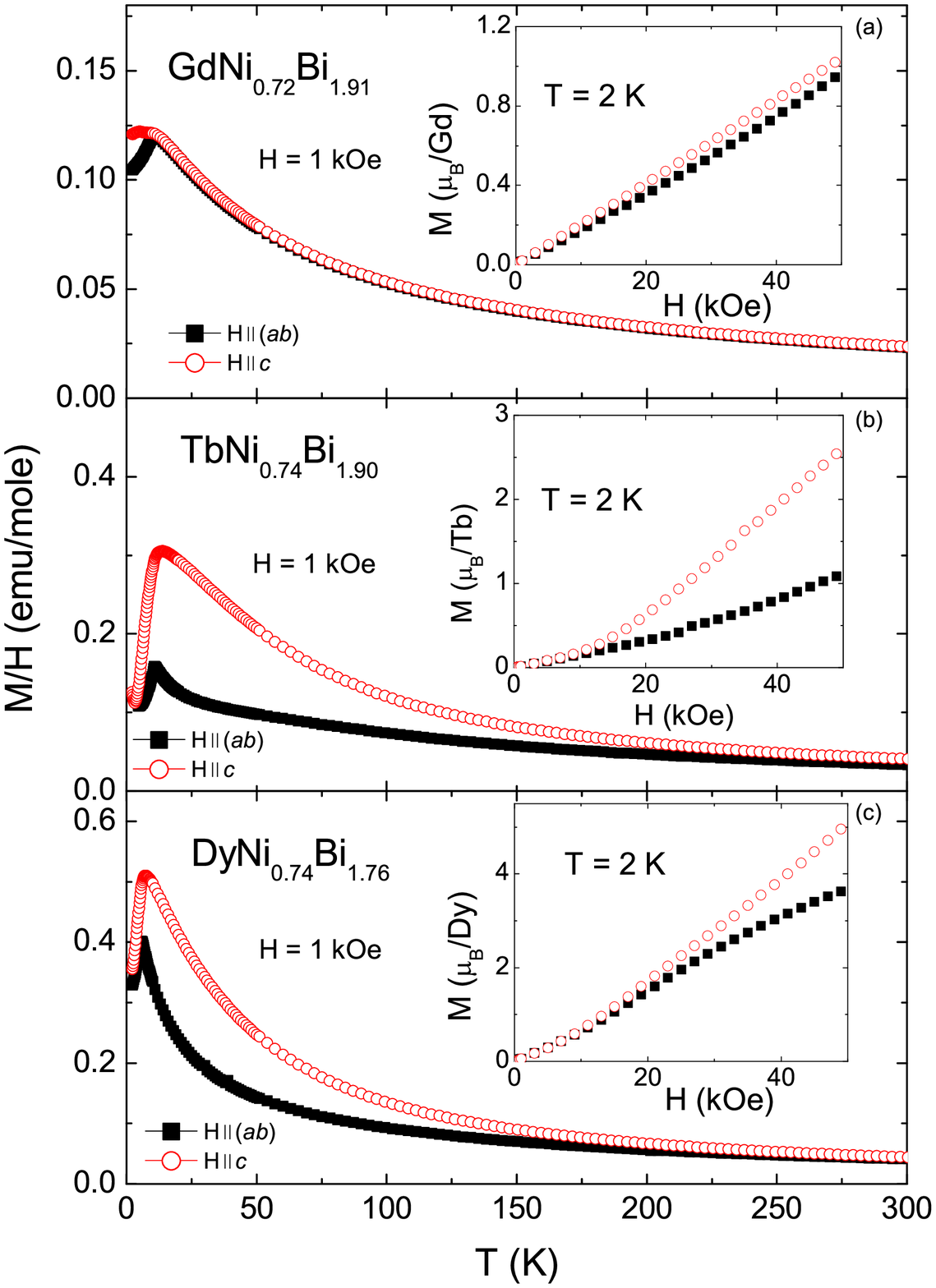}}}
\caption{(a) The anisotropic temperature-dependent magnetic susceptibility of GdNi$_{0.72}$Bi$_{1.91}$. Inset: the anisotropic magnetic isotherms of GdNi$_{0.72}$Bi$_{1.91}$.
(b) The anisotropic temperature-dependent magnetic susceptibility of TbNi$_{0.74}$Bi$_{1.90}$. Inset: the anisotropic magnetic isotherms of TbNi$_{0.74}$Bi$_{1.90}$.
(c) The anisotropic temperature-dependent magnetic susceptibility of DyNi$_{0.74}$Bi$_{1.76}$. Inset: the anisotropic magnetic isotherms of DyNi$_{0.74}$Bi$_{1.76}$.} 
\label{fig:M_all2}
\end{center}
\end{figure*}

The anisotropic temperature-dependent magnetic susceptibility data $\chi(T)=M(T)/H$ for \textit{R} = Pr, Nd, Sm, Gd--Dy measured at 1 kOe are shown in Fig. \ref{fig:M_all1} and \ref{fig:M_all2}. As can be seen in Fig. \ref{fig:M_all1} (a), PrNi$_{0.79}$Bi$_{1.98}$ manifests anisotropy with $\chi_c > \chi_{ab}$ over the whole temperature range measured. At low temperatures, sharp peaks are found at $\sim$ 6.4 K, indicating an AFM transition. At high temperatures, its magnetic susceptibility follows the modified Curie-Weiss law, giving $\theta_{ab}$ = -18 K, $\theta_{c}$ = -7 K and $\theta_{ave}$ = -12 K. The effective moment obtained from the fit of polycrystalline averaged susceptibility is $\mu_{eff}$ = 3.6 $\mu_B$ per Pr$^{3+}$ (see Table \ref{table:Temperature}), almost identical to the free ion value for Pr$^{3+}$. The anisotropic field-dependent magnetization isotherms of PrNi$_{0.79}$Bi$_{1.98}$ measured at 2 K are shown in the inset of Fig. \ref{fig:M_all1} (a). For both orientations, $M(H)$ increases as the applied field increases up to 70 kOe. For \textbf{H} $\parallel$ \textit{c}, an anomaly is seen at $\sim$ 41 kOe. For \textbf{H} $\parallel$ \textit{ab}-plane, an inflection point in $M(H)$ curve is detected at a higher field. Higher magnetic field ($H > $ 70 kOe) is needed for PrNi$_{0.79}$Bi$_{1.98}$ to be fully saturated.

For NdNi$_{0.72}$Bi$_{2.14}$, the magnetic susceptibility behaves anisotropically ($\chi_c > \chi_{ab}$) but with reduced anisotropy as compared to CeNi$_{0.80}$Bi$_{2.03}$ or PrNi$_{0.79}$Bi$_{1.98}$ (shown in Fig. \ref{fig:M_all1} (b)). At low temperatures, a sharp peak is found at $\sim$ 3.8 K, which is likely associated with an AFM transition. At high temperatures, fitting with the modified Curie-Weiss law results in $\theta_{ab}$ = -33 K, $\theta_{c}$ = -3 K and $\theta_{ave}$ = -16 K. The polycrystalline averaged susceptibility gives $\mu_{eff}$ = 3.8 $\mu_B$ per Nd$^{3+}$ (see Table \ref{table:Temperature}), which is consistent with 3.62 $\mu_B$, the expected value for Nd$^{3+}$ free ion. The anisotropic field-dependent magnetization isotherms of NdNi$_{0.72}$Bi$_{2.14}$ measured at 2 K are shown in the inset of Fig. \ref{fig:M_all1} (b). For both orientations, $M(H)$ increases as the applied field increases up to 70 kOe. A subtle inflection in the $M(H)$ curve can be observed for both orientations, indicating the possible existence of metamagnetic transitions.

The magnetic susceptibility of SmNi$_{0.78}$Bi$_{2.08}$ exhibits very subtle anisotropic behavior (Fig.\ref{fig:M_all1} (c)). An anomaly is found at $\sim$ 3.3 K (left inset of Fig.\ref{fig:M_all1} (c)). When measured with different magnetic fields, the transition temperature shows very little variance. It is probably related to an AFM ordering. Different from PrNi$_{0.79}$Bi$_{1.98}$ and NdNi$_{0.72}$Bi$_{2.14}$, the $\chi(T)$ of SmNi$_{0.78}$Bi$_{2.08}$ does not follow Curie-Weiss law but shows a tendency to saturation at high temperatures. This is commonly seen in the Sm-bearing intermetallic compounds \cite{Sergey_1999, Petrovic_2003}. This behavior may be due to Sm ion's valence fluctuation between 3+ and 2+ and/or Sm ion's excitation to upper Hund's-rule state and the associated Van Vleck paramagnetism. The field-dependent magnetization curves for SmNi$_{0.78}$Bi$_{2.08}$ show small anisotropy and increase linearly (or close to linearly) with no traces of the field-induced transitions (right inset of Fig.\ref{fig:M_all1} (c)).

The temperature-dependent magnetic susceptibility of GdNi$_{0.72}$Bi$_{1.91}$ manifests typical behavior of an antiferromagnet with no CEF effect (Fig. \ref{fig:M_all2} (a)). Due to the \textit{S}-state (\textit{L} = 0, \textit{S} = 7/2) of the Gd$^{3+}$ ion, $\chi(T)$ is virtually isotropic in the paramagnetic state. Below $T_{\rm N}$ = 9.8 K, $\chi_{ab}(T)$ decreases with decreasing temperature and $\chi_{c}(T)$ stays almost constant. These data suggest that the Gd moment orders in the basal \textit{ab}-plane. The high-temperature $\chi(T)$ follows the Curie-Weiss law with $\theta_{ave}$ = -51 K and $\mu_{eff}$ = 7.9 $\mu_B$ per Gd$^{3+}$ (see Table \ref{table:Temperature}), essentially identical to the expected value for Gd$^{3+}$ free ion. The negative sign of the paramagnetic Curie temperature suggests the presence of the antiferromagnetic correlations. The anisotropic field-dependent magnetization isotherms of GdNi$_{0.72}$Bi$_{1.91}$ measured at 2 K are shown in the inset of Fig. \ref{fig:M_all2} (a). $M(H)$ almost linearly increases as the applied field increases up to 50 kOe with a small anisotropy.

The magnetic susceptibility of TbNi$_{0.74}$Bi$_{1.90}$ is highly anisotropic (Fig. \ref{fig:M_all2} (b)), and manifests the highest ordering temperature among the members of this family with $T_{\rm N}$ $\simeq$ 10.2 K. At high temperatures, $\chi(T)$ follows the modified Curie-Weiss law, giving $\theta_{ab}$ = -109 K, $\theta_{c}$ = -19 K and $\theta_{ave}$ = -40 K. The effective moment obtained from polycrystalline averaged susceptibility is $\mu_{eff}$ = 10.0 $\mu_B$ per Tb$^{3+}$ (see Table \ref{table:Temperature}), which is consistent with 9.72 $\mu_B$, the expected value for Tb$^{3+}$ free ion. The field-dependent magnetization curve for TbNi$_{0.74}$Bi$_{1.90}$ behaves anisotropically (inset of Fig.\ref{fig:M_all2} (b)). An inflection in the $M(H)$ curve for \textbf{H} $\parallel$ \textit{c} suggests the possible existence of a metamagnetic transition.

In the case of DyNi$_{0.74}$Bi$_{1.76}$, the magnetic susceptibility is found to be anisotropic (Fig. \ref{fig:M_all2} (c)). At low temperatures, DyNi$_{0.74}$Bi$_{1.76}$ enters its AFM state at $\sim$ 5.4 K, seen by a cusp in the magnetic susceptibility curve. $\chi(T)$ at high temperatures fits the modified Curie-Weiss law, giving $\theta_{ab}$ = -33 K, $\theta_{c}$ = -12 K and $\theta_{ave}$ = -24 K. The observed effective moment from the polycrystalline averaged susceptibility is $\mu_{eff}$ = 10.7 $\mu_B$ per Dy$^{3+}$ (see Table \ref{table:Temperature}), consistent with the expected Hund's rule ground-state value, 10.65 $\mu_B$. The field-dependent magnetization curve for DyNi$_{0.745}$Bi$_{1.76}$ shows anisotropic behavior, which is shown in the inset of Fig.\ref{fig:M_all2} (c).

\section{Discussion and conclusions}

Motivated by the recent claims about CeNi$_{0.8}$Bi$_{2}$ \cite{Hosono1, Hosono2, Hosono3} and previous studies of rare-earth compounds \cite{Myers_1999, Sergey_1999, Petrovic_2003}, we have synthesized single crystalline \textit{R}Ni$_{1-x}$Bi$_{2 \pm y}$ (\textit{R} = La--Nd, Sm, Gd--Dy) samples by using Bi as a flux. Detailed resistivity, magnetic susceptibility and magnetization measurements were performed to study the properties of \textit{R}Ni$_{1-x}$Bi$_{2 \pm y}$. The crystals form as plates, and can be identified as having a \textit{P4/nmm} structure. We have determined the Ni and Bi concentrations and seen clear evidence of the associated, disorder scattering manifest in the low RRR values. 

Superconducting features have been observed in the transport measurements for LaNi$_{0.84}$Bi$_{2.04}$ and CeNi$_{0.80}$Bi$_{2.03}$, as well as the other members in this family. However, the transition temperatures coincide with the $T_{\rm c}$ of film Bi, NiBi and/or NiBi$_{3}$, and the features are irreproducible and sample-dependent. Moreover, for LaNi$_{0.84}$Bi$_{2.04}$, both of the ZFC and FC superconducting volume fractions at 50 Oe are $< 3\%$. The low-field magnetic susceptibility of CeNi$_{0.80}$Bi$_{2.03}$ is positive. All these strongly suggest that the superconductivity in \textit{R}Ni$_{1-x}$Bi$_{2 \pm y}$ (\textit{R} = La--Nd, Gd--Dy) is due to minority, second phases. 

\begin{figure}
\begin{center}
\resizebox*{7.5cm}{!}{\includegraphics{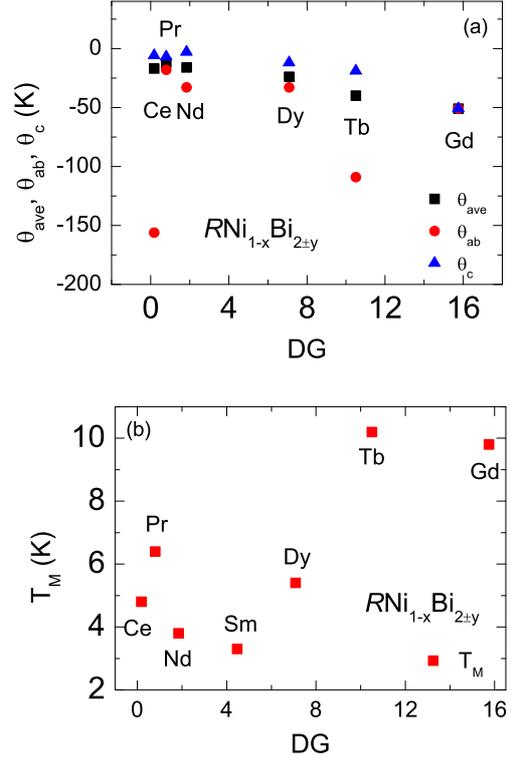}}%
\caption{ Changes of (a) paramagnetic Curie temperature $\theta_{ave}$, $\theta_{ab}$, $\theta_{c}$ and (b) magnetic ordering temperatures with the de Gennes parameter DG.}%
\label{fig:DG}
\end{center}
\end{figure}

The high-temperature magnetic susceptibilities of \textit{R}Ni$_{1-x}$Bi$_{2 \pm y}$ (\textit{R} = Ce--Nd, Gd--Dy) show local-moment like behaviors. For the whole \textit{R}Ni$_{1-x}$Bi$_{2 \pm y}$ family, Ni is non-moment bearing. The values of the effective magnetic moment in the paramagnetic state are close to the theoretical values of the trivalent rare earth ion. $\mu_{eff}$, $\theta_{ab}$, $\theta_{c}$ and $\theta_{ave}$ values, obtained by fitting with the modified Curie-Weiss law are summarized in Table \ref{table:Temperature}. The local-moment ordering is likely governed by the indirect exchange interactions between the rare earth ions mediated by the conduction electrons (RKKY interaction). The negative sign of the paramagnetic Curie temperatures $\theta_{ab}$, $\theta_{c}$ and $\theta_{ave}$ indicates the dominate interactions in this system are antiferromagnetic. Based on the Weiss molecular field theory, both $\theta_{ave}$ and the magnetic ordering temperature $T_{\rm M}$ are expected to be proportional to the de Gennes factor DG = $(\mathtt{g}_J -1)^2J(J+1)$. Here $\mathtt{g}_J$ is the Land$\acute{\rm e} $ $\mathtt{g}$ factor and \textit{J} is the total spin angular momentum \cite{DG_1962}. As shown in Fig. \ref{fig:DG}, by removing the CEF effect, the paramagnetic Curie temperature $\theta_{ave}$ follows the DG scaling quite well except for \textit{R} = Ce. This deviation is probably related to the hybridization between 4\textit{f} and the conduction electrons. However, significant deviations from linearity are present for the scaling of the magnetic ordering temperatures $T_{\rm M}$. This can occur when a strong CEF constrains the moments to either along the \textit{c}-axis or within the basal plane \cite{DG_1982}. This may be responsible for the higher value of TbNi$_{0.74}$Bi$_{1.90}$, as it shows strong anisotropy in Fig. \ref{fig:M_all2} (b). The decreasing unit cell volume in this series (lanthanide contraction) may lead to changes in the conduction electron density and/or the exchange constants (due to shorter atomic distances), and these changes could also be responsible for this deviation from the de Gennes scaling. Similar to their isostructural compounds \textit{R}AgSb$_{2}$ \cite{Myers_1999} and \textit{R}AgBi$_{2}$ \cite{Petrovic_2003}, most of the \textit{R}Ni$_{1-x}$Bi$_{2 \pm y}$ members manifest antiferromagnetic ordering at low temperatures. \textit{R}Ni$_{1-x}$Bi$_{2 \pm y}$ shows more anisotropy in their magnetic properties than the \textit{R}AgBi$_{2}$ series, but less than the \textit{R}AgSb$_{2}$ family's. $\theta_{ave}/T{_M}$ values characterizing the level of frustration \cite{Frustration} for \textit{R}Ni$_{1-x}$Bi$_{2 \pm y}$ although enhanced, between 1.9 and 5.2, are not significantly different from those found for either the \textit{R}AgBi$_{2}$ or the \textit{R}AgSb$_{2}$ compounds, which means the values are not clearly related to the site disorder.

\section*{Acknowledgements}
This work was carried out at the Iowa State University and supported by the AFOSR-MURI grant No. FA9550-09-1-0603 (X. Lin and P. C. Canfield). S. L. Bud'ko was supported by the U.S. Department of Energy, Office of Basic Energy Science, Division of Materials Sciences and Engineering. Part of this work was performed at Ames Laboratory, US DOE, under Contract No. DE-AC02-07CH11358.

\begin{table*}[ht]
\caption{WDS elemental analysis (in atomic $\%$) for \textit{R}Ni$_{1-x}$Bi$_{2 \pm y}$ single crystals.} 
\centering
\setlength{\tabcolsep}{2.5pt}
\begin{tabular}{c c c c c c c c c c c} 
\hline\hline 
Compound &&	\textit{N} (number of points analyzed) && \textit{R} && Ni && Bi && Stoichiometry (WDS) \\ [0.5ex]
\hline
La	&&	12	&&	25.74(14)	&&	21.65(6)	&&	52.61(17)	&&	LaNi$_{0.84}$Bi$_{2.04}$\\		
Ce	&&	12	&&	26.09(11)	&&	20.99(14)	&&	52.92(21)	&&	CeNi$_{0.80}$Bi$_{2.03}$\\		
Pr	&&	11	&&	26.49(8)	&&	21.06(11)	&&	52.45(14)	&&	PrNi$_{0.79}$Bi$_{1.98}$\\		
Nd	&&	12	&&	25.91(11)	&&	18.73(4)	&&	55.36(12)	&&	NdNi$_{0.72}$Bi$_{2.14}$\\		
Sm	&&	12	&&	25.96(10)	&&	20.13(6)	&&	53.91(13)	&&	SmNi$_{0.78}$Bi$_{2.08}$\\		
Gd	&&	11	&&	27.53(11)	&&	19.88(16)	&&	52.58(24)	&&	GdNi$_{0.72}$Bi$_{1.91}$\\		
Tb	&&	12	&&	27.51(6)	&&	20.36(7)	&&	52.13(11)	&&	TbNi$_{0.74}$Bi$_{1.90}$\\		
Dy	&&	11	&&	28.53(21)	&&	21.14(10)	&&	50.33(30)	&&	DyNi$_{0.74}$Bi$_{1.76}$\\		
[1ex]
\hline
\end{tabular}
\label{table:WDS}
\end{table*}

\begin{table*}[ht]
\caption{Refined unit cell parameters from powder x-ray diffraction for \textit{R}Ni$_{1-x}$Bi$_{2 \pm y}$ compounds.} 
\centering
\begin{tabular}{c c c c c}
\hline\hline
Compound & \textit{a} (\r{A}) & \textit{c} (\r{A}) & \textit{V} ({\r{A}}$^3$) \\ [0.5ex]
\hline
La	&	4.56(1)	&	9.78(2)	&	204.02(3)	\\
Ce	&	4.54(1)	&	9.64(2)	&	199.57(3)	\\
Pr	&	4.52(1)	&	9.60(2)	&	196.68(3)	\\
Nd	&	4.52(1)	&	9.53(2)	&	194.96(3)	\\
Sm	&	4.50(1)	&	9.42(2)	&	191.27(3)	\\
Gd	&	4.49(1)	&	9.37(2)	&	189.20(3)	\\
Tb	&	4.48(1)	&	9.30(2)	&	187.52(3)	\\
Dy	&	4.47(1)	&	9.29(2)	&	185.82(3)	\\
[1ex]
\hline
\end{tabular}
\label{table:powder}
\end{table*}

\begin{table*}[ht]
\caption{Magnetic ordering temperatures, anisotropic Curie temperatures and effective magnetic moment in paramagnetic state for \textit{R}Ni$_{1-x}$Bi$_{2 \pm y}$.} 
\centering
\setlength{\tabcolsep}{2.5pt}
\begin{tabular}{c c c c c c c} 
\hline\hline 
Compound&$\theta_{ab}$ (K)&$\theta_{c}$ (K)&$\theta_{ave}$ (K)&$\mu_{eff}$ ($\mu_B$)&$\chi_0$ (10$^{-4}$ emu/mole) &$T_{\rm M}$ (K)  \\ [0.5ex]
\hline
Ce 	&	-156	&	-6	&	-17	&	2.4	&	10	&	4.8	\\
Pr	&	-18	&	-7	&	-12	&	3.6	&	1	&	6.4	\\
Nd	&	-33	&	-3	&	-16	&	3.8	&	2	&	3.8	\\
Sm	&		&	 	&	 	&	 	&		&	3.3	\\
Gd	&	-51	&	-51	&	-51	&	7.9	&	10	&	9.8	\\
Tb	&	-109	&	-19	&	-40	&	10	&	-12	&	10.2	\\
Dy	&	-33	&	-12	&	-24	&	10.7	&	27	&	5.4	\\
[1ex]
\hline
\end{tabular}
\label{table:Temperature}
\end{table*}

\end{document}